\documentclass[final,english]{bullsrsl}[2022/06/15]

\usepackage[latin1]{inputenc}
\usepackage[T1]{fontenc}

\usepackage{natbib} 
\usepackage{graphicx}
\usepackage{floatrow}

\newcommand{\aap}{A\&A}
\newcommand{\apj}{ApJ}
\newcommand{\apjl}{ApJL}

\newcommand{\mnras}{MNRAS}
\newcommand{\solphys}{SoPh}
\newcommand{\ssr}{SSRv}
\defcitealias{2023MNRAS.tmp.2323R}{RG23}

\newcommand{\amp}{\&}

\begin{document}
\title{Exploring magnetic coupling of solar atmosphere through frequency modulations of 3-min slow magnetoacoustic waves }

\author[affil={1,2}]{Ananya}{Rawat}
\author[affil={1},corresponding]{Girjesh}{Gupta}    

\affiliation[1]{Udaipur Solar Observatory, Physical Research Laboratory, Udaipur, India 313001}
\affiliation[2]{Department of Physics, Indian Institute of Technology Gandhinagar, India 382355}
\correspondance{girjesh@prl.res.in; ananyarawat@prl.res.in}
\date{\today}
\maketitle


%

\begin{abstract}
Coronal fan loops rooted in sunspot umbra show outward propagating waves with subsonic phase speed and period around 3-min. However, their source region in the lower atmosphere is still ambiguous. We performed  multi-wavelength observations of a clean fan loop system rooted in sunspot observed by Interface Region Imaging Spectrograph (IRIS) and Solar Dynamics Observatory (SDO). We utilised less explored property of frequency modulation of these 3-min waves from the photosphere to corona, and found them to be periodic with the ranges in 14-20 min, and 24-35 min.  Based on our findings, we interpret that 3-min slow waves observed in the coronal fan loops are driven by 3-min oscillations observed at the photospheric footpoints of these fan loops in the umbral region. We also explored any connection between 3-min and 5-min oscillations observed at the photosphere, and found them to be poorly understood. Results provide clear evidence of magnetic coupling of the solar umbral atmosphere through propagation of 3-min waves along the fan loops at different atmospheric heights. 
\end{abstract}


\section{Introduction}

Magnetohydrodynamic (MHD) waves and oscillations are ubiquitous in the solar atmosphere. Propagation properties of these waves were first reported by \citet{1997ApJ...476..357O} in polar coronal holes, and by \citet{1999Sci...285..862N} and \citet{2002A&A...387L..13D} in coronal loops. Comprehensive details on their properties can be found  in \citet{2015LRSP...12....6K}, \citet{2011SSRv..158..267B,2021SSRv..217...76B}, and \citet{2023LRSP...20....1J} for sunspot regions, coronal hole regions, and lower atmosphere, respectively to name a few. These waves carry sufficient energy, and can be an efficient mechanism for coronal heating. Recently, lot of studies have been dedicated to find observational evidence of wave damping while propagating along various structures in the solar atmosphere \citep[e.g.][]{2012ApJ...753...36H,2014ApJ...784...29M,2014A&A...568A..96G,2014ApJ...789..118K,2017ApJ...836....4G,2018NatPh..14..480G,2019A&A...627A..62G}. Details and current status of the wave heating mechanism of solar corona can be found in recent reviews such as \citet{2012RSPTA.370.3193D}, \citet{2020SSRv..216..140V} etc. Although there exist now ample reports on propagation and damping characteristics of waves in the different layers of the solar atmosphere, observational reports on their generation and source regions are still rare.

In the lower atmosphere, sunspot regions generally show strong power in  5-min oscillation bands and significant power in 3-min oscillation bands at the photosphere \citep{2000ApJ...534..989B}. However at heights above the temperature minimum, 3-min oscillations dominate over 5-min oscillations \citep[e.g.,][]{2016PhDT........15L}. In the chromosphere, flashes are found at random locations in the umbra with a period of about 3-min \citep{1969SoPh....7..351B} which are often interpreted as signatures of upward propagating magnetoacoustic shock waves \citep[see details in][]{2006ApJ...640.1153C}. 

In the corona, fan loops rooted in sunspot umbra show continuous presence of propagating disturbances  with period around 3-min \citep[e.g.][]{2002SoPh..209...61D}. Similar propagating disturbances with the period around 15-min are also observed along structures in coronal hole \citep[e.g.][]{1998ApJ...501L.217D,2010ApJ...718...11G}. These propagating disturbances have subsonic speeds and are often interpreted as propagating slow magnetoacoustic waves \citep[e.g.][]{2012SoPh..279..427K,2012A&A...546A..93G}. These 3-min slow magnetoacoustic waves also show modulations in their amplitude with periods in the range of 20-30 min \citep{2020A&A...638A...6S}. These waves also show evidence of modulations in frequency with time while propagating \citep{2012A&A...539A..23S}.  Although these waves are ubiquitous in the different coronal structures, observational evidence of their source region is still very rare \citep[e.g.][]{2012ApJ...757..160J,2015ApJ...812L..15K}. 
 
In the sunspots, umbral flashes influence the propagation properties of different sunspot waves as noted by \citet{2017ApJ...850..206S}.  \citet{2012ApJ...757..160J} noted that coronal fan loops are rooted in the umbral dots at the photosphere which showed enhanced power of 3-min oscillation.  \citet{2015ApJ...812L..15K} utilised the properties of amplitude modulation of 3-min oscillations at different atmospheric layers, and associated presence of 3-min slow magnetoacoustic waves in the corona with the 5-min oscillations of solar p-mode at the photosphere.  On the other hand, \citet{2013A&A...554A.146K} and \citet{2017ApJ...836...18C} found no photospheric and coronal connections. This leads to a mixed opinion on the origin of these 3-min waves found in the umbral atmosphere. Therefore, recently 
\citet[][henceforth referred to as \citetalias{2023MNRAS.tmp.2323R}]{2023MNRAS.tmp.2323R} carried out a detailed investigation to identify the source region of 3-min waves and oscillations observed in the upper atmosphere. They utilised amplitude modulation property of 3-min waves and provided a very comprehensive understanding of their source region at the photosphere.

\begin{figure}[th]
    \includegraphics[width=0.95\textwidth]{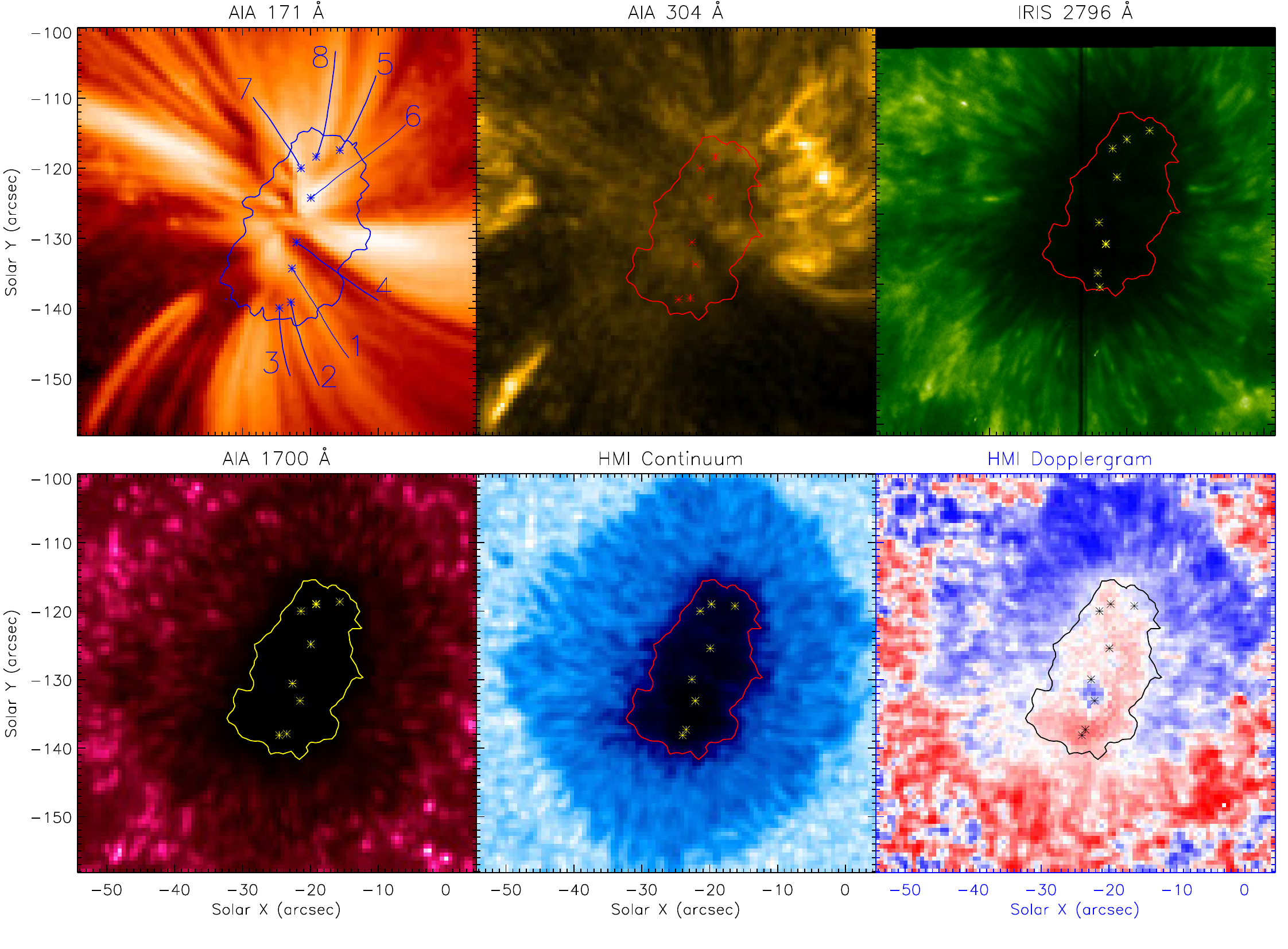}
    \caption{Images of sunspot observed from different AIA, IRIS, and HMI passbands as labeled. Fan loop system is observed in AIA 171 \AA\ passband. Identified loop locations are marked with asterisks (*). Traced coronal loops are also drawn in AIA 171 \AA\ passband for visualisation purpose only. Contours indicate the umbra-penumbra boundary as obtained from the HMI continuum image.}
    \label{fig:map}
\end{figure}

 In this work, we present a multi-wavelength analysis of propagation of 3-min slow magnetoacoustic waves from the photosphere to the corona along fan loop structures rooted within a sunspot umbra. Coronal fan loops are steady, quiescent, and long loops whose thermal properties are described in \citet{Ghosh_2017}. Our motivation is to probe the source region of these waves at the photosphere, and thus the magnetic connectivity of the solar atmosphere by utilising less explored frequency modulation properties of 3-min waves which is in addition to results provided by \citetalias{2023MNRAS.tmp.2323R}. We present details of observation in Section~\ref{sec:obs}, data analysis and results in Section~\ref{sec:analysis}, and discussion and summary in Section~\ref{sec:discussion}.

\section{Observations}
\label{sec:obs}

To investigate the source region of waves in coronal fan loops, and thus magnetic connectivity of the solar atmosphere, we utilize multi-wavelength observations of a sunspot belonging to Active region NOAA 12553 for a duration of 4 hours on June 16, 2016 starting from 07:19:13 UT. Sunspot was observed by Atmospheric Imaging Assembly \citep[AIA;][]{2012SoPh..275...17L}, Helioseismic and Magnetic Imager \citep[HMI;][]{2012SoPh..275..207S} both onboard Solar Dynamics Observatory \citep[SDO;][]{2012SoPh..275....3P}, and Interface Region Imaging Spectrograph \citep[IRIS;][] {2014SoPh..289.2733D}. Sunspot region is shown in Figure~\ref{fig:map} for all the analysed passbands as labeled. Details on data preparation are described in \citetalias{2023MNRAS.tmp.2323R}.

\section{Data Analysis and Results}
\label{sec:analysis}

Figure~\ref{fig:map} shows analysed fan loop structures in AIA 171 \AA\ passband.  Asterisks (*) represent the coronal foot-points of fan loops in the AIA 171 \AA\ image. Locations from where the loops appear to start in coronal AIA 171 \AA\ image are considered as loop foot-points in the corona. These locations are used as a reference for identifying locations of these loops in the lower solar atmosphere \citepalias[details on methodology is provided in][]{2023MNRAS.tmp.2323R}. Identified loop locations in the lower atmosphere are marked with asterisks. We have identified eight fan loops emanating from the sunspot umbra in our study and labeled them accordingly. Associated coronal loops are also drawn for visualisation purpose only. In this work, we analyse all the identified fan loop locations. However, we present results only from loop foot-point 6 as a representative example. Results from all the other loop foot-points are summarized in Tables~\ref{tab:ama} and \ref{tab:ama5}.

Although 3-min oscillations are observed at all the layers of upper sunspot atmosphere \citep{2015LRSP...12....6K}, they are sometimes detected at the photospheric heights also though weaker than  5-min period oscillations  \citep{2000ApJ...534..989B}. Since we have found significant power in the 3-min period band at each atmospheric heights studied here \citepalias[see details in][]{2023MNRAS.tmp.2323R}, we are utilising it for detailed investigation. We observe oscillations in the form of unclean wave packets at the coronal foot-point of loop 6 as shown in the top panel of Figure~\ref{fig:wavelet}. This is due to the several nearby power peaks present in the period band 2-3.8 min as explained in \citetalias{2023MNRAS.tmp.2323R}.

\subsection{Wavelet analysis}
 
\begin{figure*}[th]
    \centering
    \includegraphics[width=0.7\textwidth,angle=90]{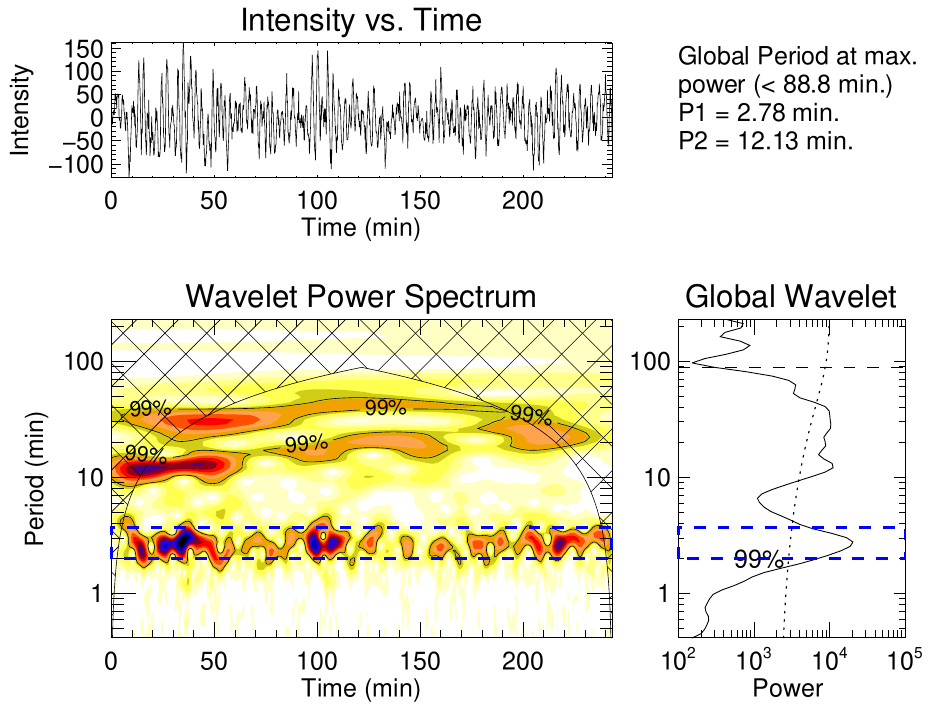}
    \caption{The top panel shows the 16 min background subtracted light curve as obtained at the coronal foot-point of loop 6 from AIA 171 \AA\ passband. Bottom left panel shows the colour-coded wavelet power spectrum of the background subtracted light curve with 99\% confidence-level contours. The time axis starts at 7:19 UT. Different colour contours show varying power densities where blue indicates the highest. Region marked with crossed lines denote cone-of-influence. The bottom right panel shows the global wavelet power spectrum. The thin black dashed line indicates the maximum detectable period. The dotted line specifies 99\% confidence level curve. The periods P1 and P2 are the locations of the first two maxima in the global wavelet spectrum. The overplotted blue colour box indicates the period window extracted to obtain detailed property of 3-min oscillations.}
    \label{fig:wavelet}
\end{figure*}

In the top panel of Figure~\ref{fig:wavelet}, we plot 16 min background subtracted light curve of loop 6 foot-point as obtained from AIA 171 \AA\ passband. 16 min background was obtained by applying 80-point (16 min) running average to the light curve under study. In the bottom left panel of Figure~\ref{fig:wavelet}, we show the wavelet power spectrum of same background subtracted light curve. To obtain the wavelet power spectrum, we are utilising a tool developed in IDL by \citet{1998BAMS...79...61T}. The left panel shows the wavelet power spectrum with time on the x-axis and period on the y-axis. Panel clearly shows variation of oscillatory power with time. Here, the region marked with crossed lines is called the cone-of-influence (COI) which refers to the region where the transform suffers from the edge effects. Oscillation periods in this region are unreliable. In the right panel of Figure~\ref{fig:wavelet}, we show the global wavelet power spectrum, which is obtained by taking the average over the time domain of the wavelet transform. The thin dashed line shows the maximum period detectable from the wavelet analysis due to the COI. The dotted line shows 99$\%$ confidence level curve above which the power is considered reliable. 

We extract 2-3.8 min period window from the wavelet power spectrum as shown by the blue colour box in the bottom panel of Figure~\ref{fig:wavelet} to obtain more detailed properties of 3-min oscillations. We obtain frequency modulations of 3-min waves from this period window as describe in the following subsection.

\subsection{Frequency modulations}

\begin{figure*}[th]
    \centering
    \includegraphics[width=0.88\textwidth]{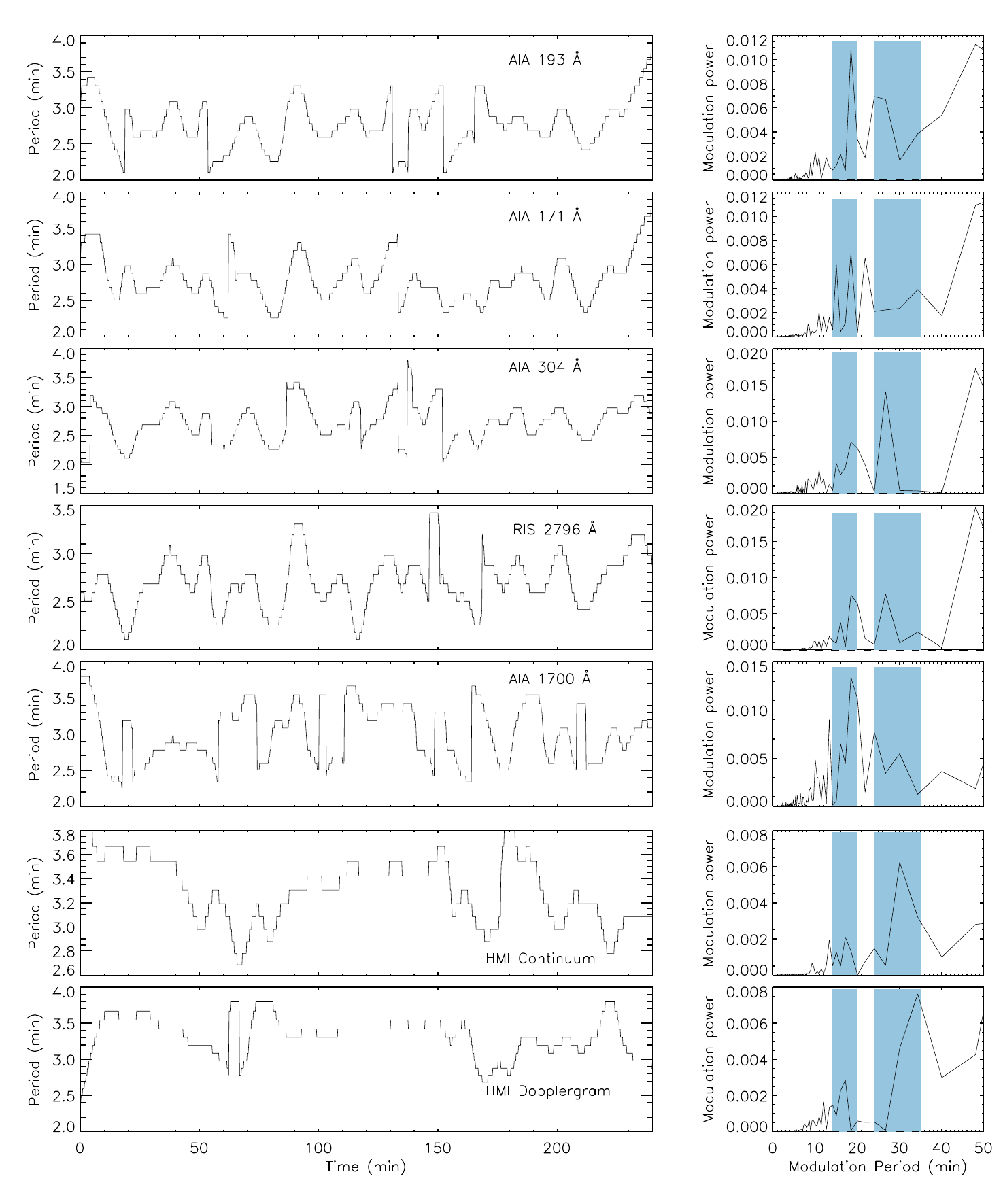}
    \caption{Left panels show frequency modulations of 3-min oscillations extracted from the wavelet spectrum for different passbands as labeled. Right panels show the FFT power spectrum of corresponding frequency modulation curves. Shaded regions in sky blue colour highlight dominant modulation periods observed at different atmospheric heights.}
    \label{fig:fm}
\end{figure*}

To determine the frequency modulation of 3-min waves, we first obtained the wavelet spectrum of background subtracted light curves for all the passbands. We then smoothened the wavelet spectrum by applying a running average of 7-point on time and 3-point on period axes. From this smoothened wavelet spectrum, we extracted 2-3.8 min period window. With this subset, we further extract the period at which power is maximum at each time frame. Following the same procedure, we obtained period or frequency variation of 3-min oscillation with time at all the atmospheric heights. Obtained period variations with time (i.e. frequency modulations) are plotted in the left panels of Figure~\ref{fig:fm} for all the passbands as labeled.

We also obtained the FFT power spectrum of frequency modulations and plotted it in the right panels of Figure~\ref{fig:fm}. Here, we can see that dominant 3-min frequency modulation periods are approximately in the range 14-20 min and 24-35 min  which are present at all the layers of the solar atmosphere. These modulation periods are shown as shaded regions in the right panel of Figure~\ref{fig:fm}. These similarity in modulation periods signify that 3-min oscillations are essentially coupled together at different atmospheric layers. Thus, provides clear evidence of upward propagating waves from the photosphere to the corona along the observed fan loop locations. We also noticed that power peaks within 24-35 min shaded region are shifting towards the left for coronal heights. However, these shifts are within the error range and demands a high frequency resolution data to reduce any such shifts in modulation periods.    
  
\begin{figure*}[th]
    \centering
    \includegraphics[width=0.85\textwidth]{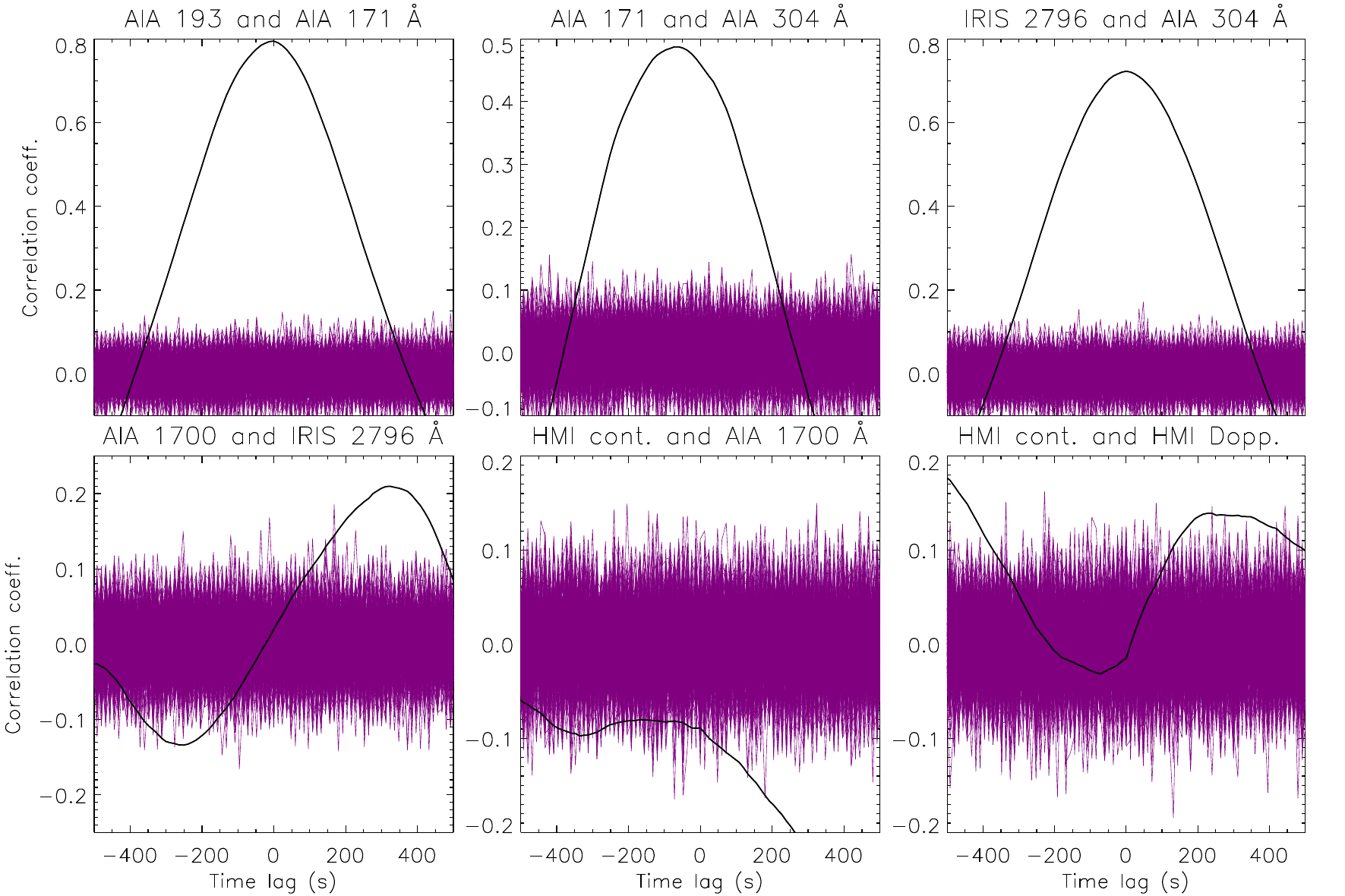}
    \caption{Correlation coefficients at different time lags obtained between frequency modulation curves for different atmospheric heights as shown in the left panel of Figure~\ref{fig:fm}. Purple colour lines are obtained by carrying out correlation analysis on randomized frequency modulation curves.}
    \label{fig:co_fm}
\end{figure*}

We also cross-correlated the frequency modulation curves obtained at different atmospheric heights with the nearest atmospheric layers. These correlation curves are plotted in Figure~\ref{fig:co_fm}. Correlations between different height pairs follow the similar pattern as obtained from amplitude modulations in \citetalias{2023MNRAS.tmp.2323R}, but correlation values are smaller. Especially, for pairs of AIA 1700 \AA\ and HMI continuum, and HMI continuum and Dopplergram,  correlations are poor and within the error range whereas for IRIS 2796 and AIA 1700 \AA\ pair, correlation is weak but crosses the error range. Therefore, correlation analysis between frequency modulation curves provide evidence of magnetic connectivity of the solar atmosphere from temperature minimum region (AIA 1700 \AA )  to corona (AIA 171 and 193 \AA ). However, all the frequency modulation curves show similar modulation periods. Together these findings provide clear evidence of magnetic connectivity of the whole solar atmosphere. Further to check the reliability of correlation coefficient values, we performed randomization boot-strap analysis \citepalias[see details in][]{2023MNRAS.tmp.2323R}. In Figure~\ref{fig:co_fm}, we clearly see that correlation values are either above or below the error range provided by boot-strap analysis. We also noticed that correlation between modulation curves decreases as we move into the lower atmosphere but values are still outside the error range. This implies that  modulations from upper heights are correlated to modulations from lower heights. Henceforth, we can clearly interpret that 3-min oscillations and thus frequency modulations originate at the photosphere, and are moving upward as waves move upward.

\subsubsection{Relation between 3-min and 5-min oscillations at the photosphere}

\begin{figure*}[th]
    \centering
    \includegraphics[width=0.57\textwidth]{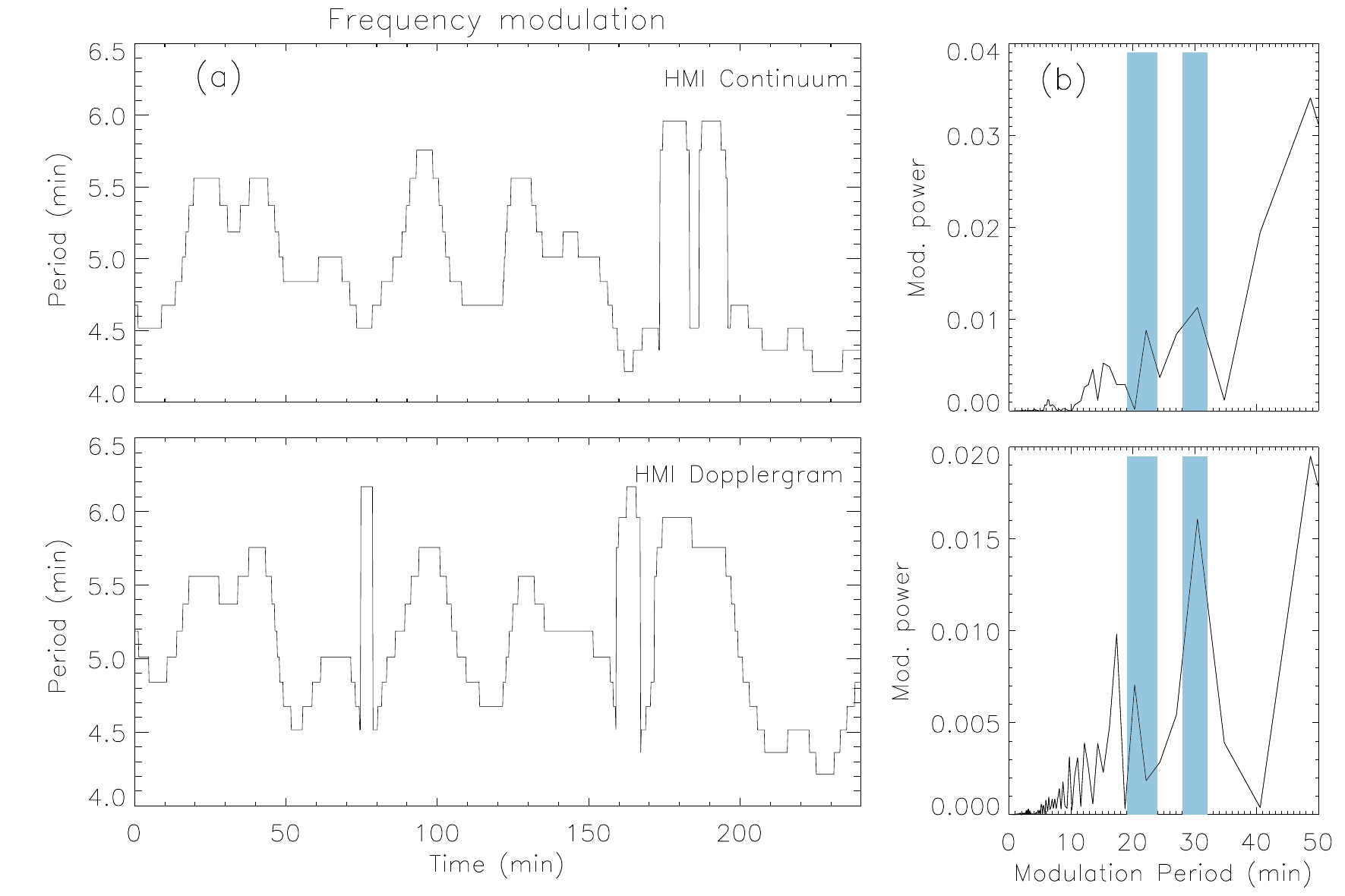} \includegraphics[width=0.42\textwidth]{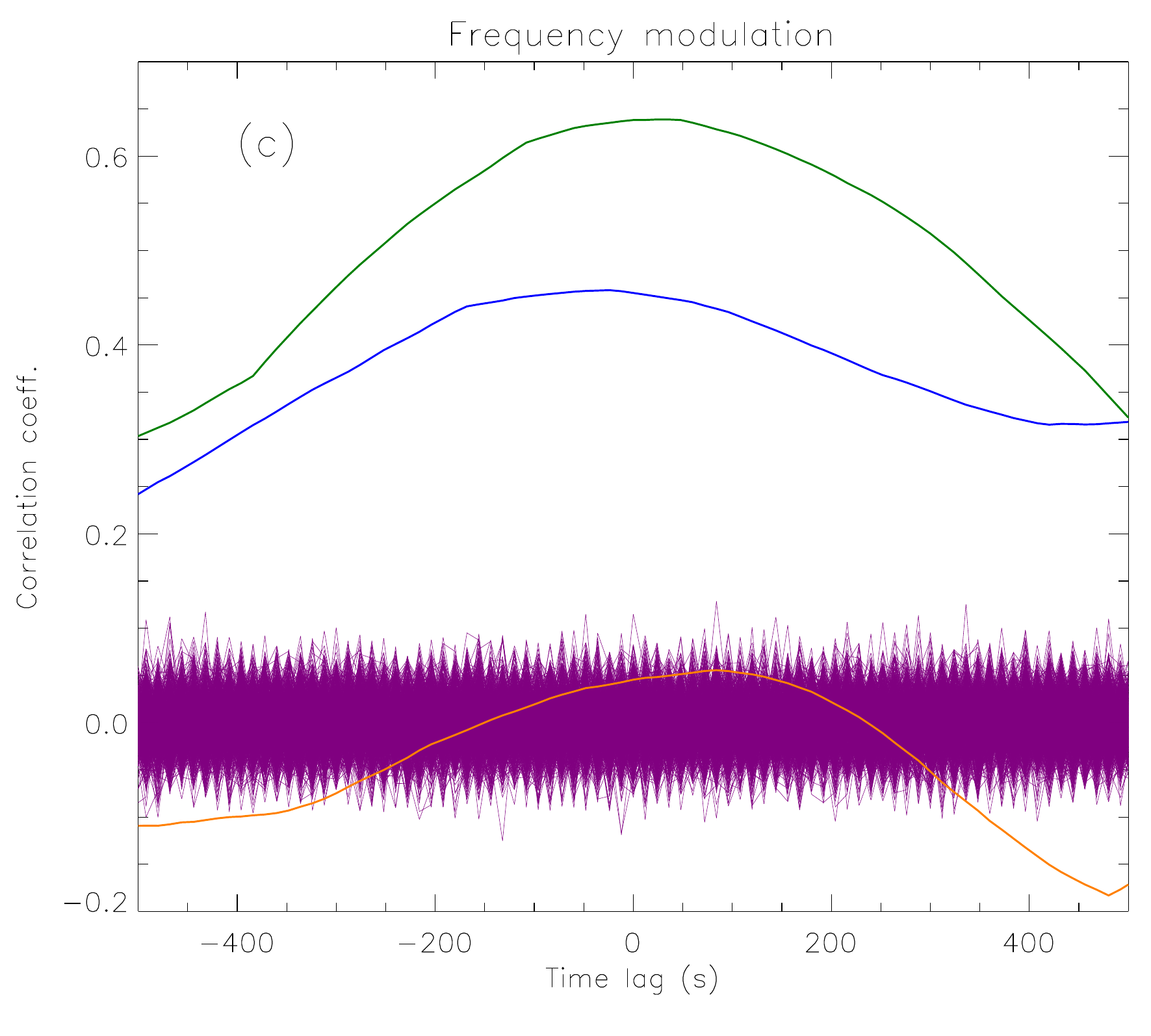}
    \caption{(a) Frequency modulations of 5-min oscillations at the photosphere obtained from HMI continuum and Dopplergram as labeled. (b) The FFT power spectrum of corresponding modulations. Shaded regions in sky blue colour highlight dominant modulation periods observed in different HMI passbands. (c) Correlations with respect to time lags obtained between frequency modulations of 3-min and 5-min oscillations. Green, blue, and orange colour lines show correlations obtained between frequency modulation curves of 5-min oscillations from HMI continuum and Dopplergram, modulations of 3-min and 5-min oscillations from HMI continuum, and modulations of 3-min and 5-min oscillations from HMI Dopplergram respectively. Purple colour lines are obtained by carrying out correlation analysis on respective randomized modulation curves.}
    \label{fig:fm5}
\end{figure*}

Since both 3-min and 5-min oscillations are observed at the photosphere \citep[e.g.,][]{2006ApJ...640.1153C,2023MNRAS.tmp.2323R}, we also obtained frequency modulations of 5-min oscillation at the photosphere so as to check any relation between both the oscillations. For this, we again performed similar wavelet analysis by choosing the period window between 4.1-6.2 min. In Figure~\ref{fig:fm5}, we plot frequency modulations of 5-min oscillations as labeled (panel a), and their respective FFT power spectrum (panel b). From the plots, we clearly see power peaks between 19-24 min, and 28-32 min periods which are showed as shaded regions from both HMI continuum and Dopplergram as labeled. We also notice
 a weak modulation period of around 16-17 min in both the passbands. 

Correlation between modulations from both the HMI passbands for 5-min oscillations is about 0.64 with time lag of $\approx 24$ s as shown with the green line in Figure~\ref{fig:fm5}(c). On comparing modulation periods of both 3-min and 5-min oscillations from Figures~\ref{fig:fm} and \ref{fig:fm5}(b), we find that longer modulation periods above 25 min and weaker modulation period around 16-17 min are present in both the oscillations. Therefore, we further carry out correlation analysis to verify any coupling between these modulations, and thus any coupling between 3-min and 5-min oscillations.

In Figure~\ref{fig:fm5}(c), we plot correlation coefficients with respect to time lags obtained between frequency modulations of 3-min and 5-min oscillations observed from HMI continuum and Dopplergram as labeled. Maximum correlation coefficients between modulations of 3-min and 5-min oscillations from HMI continuum data is about 0.46 with a time lag of $\approx-24$ s whereas corresponding correlations obtained from HMI Dopplergram data is very small and within the error range. Here again, error range is obtained from randomization boot-strap analysis as before. Therefore, these correlation values indicate poor or weak connection between 3-min and 5-min oscillations at the photosphere. Henceforth, any connection between them at the photosphere is still unclear as also noticed by \citetalias{2023MNRAS.tmp.2323R}, and thus demands a dedicated study to explore their connection in detail.

We also performed a similar analysis on other fan loops identified in Figure~\ref{fig:map}. Summary of frequency modulation periods obtained for 3-min and 5-min oscillations at various atmospheric heights for all the eight loops are provided in Tables~\ref{tab:ama} and \ref{tab:ama5}, respectively. Obtained results suggest more or less similar statistics as presented by \citetalias{2023MNRAS.tmp.2323R} from amplitude modulations of 3-min and 5-min oscillations for all the loops at various atmospheric heights. On comparison, we find that periods of amplitude and frequency modulations show similar modulation periods with at least one common modulation period observed at all atmospheric heights. Results indicate that 3-min waves observed in coronal fan loops originate at the photosphere, and are propagating upward based on presence of at least one common modulation period at all atmospheric height \citep[e.g.,][]{2015ApJ...812L..15K,2023MNRAS.tmp.2323R}. We also find several common modulation periods between amplitude and frequency modulations at all atmospheric heights which implies that these modulations are also connected to each other. Moreover, we also noted strong anti-correlations between amplitude and frequency modulations at chromospheric heights (AIA 1700 \AA\ or IRIS 2796 \AA ) for all the loops. Results from all the loops again provide clear evidence of magnetic connectivity of the whole solar atmosphere. 

\begin{table*}
    \caption{Frequency modulation periods (min) of 3-min oscillations observed in various fan loop locations at different atmospheric heights as marked in Figure~\ref{fig:map}.}
     \label{tab:ama}
   \centering
   \small
    \begin{tabular}{|c|r|r|r|r|r|r|r|r|} 
    \hline
Passbands (\AA)  & loop 1 & loop 2 & loop 3 & loop 4 & loop 5 & loop 6 & loop 7 & loop 8 \\
\hline
AIA 193        &14, 34  & 17, 34 & 15, 30   &17, 27  & 18, 40 &19, 26  &19, 34   & 15, 34  \\ \hline
AIA 171        &15, 27  & 18, 34 & 15, 34   & 15    & 40    & 19, 34 & 20, 34  & 15, 34\\ \hline
AIA 304        &20, 34  & 18, 30 & 15, 34   & 16    & 18, 40 & 18, 27 & 19, 30  & 34 \\ \hline
IRIS 2796      &13, 30  & 20, 30 & 18, 34   & 16, 28 & 40    & 18, 27 & 19, 30  & 16, 34\\ \hline
AIA 1700       &16, 30  & 17, 30 & 17      & 14, 30 & 19, 40 & 19, 30 & 20, 30  & 24, 30\\ \hline
HMI continuum  &17, 34  & 34    & 18, 35   & 13, 34 & 14, 30 & 17, 30 & 24, 40  & 20, 34\\ \hline
HMI Dopplergram &20, 30 & 27, 40 & 27      & 13, 30 & 14, 30 & 17, 34 & 20, 34  & 20, 40\\ \hline
    \end{tabular}
\end{table*}

\begin{table*}
    \caption{Frequency modulation periods (min) of 5-min oscillations observed in various fan loop locations at the photosphere as marked in Figure~\ref{fig:map}.}
     \label{tab:ama5}
   \centering
   \small
    \begin{tabular}{|c|r|r|r|r|r|r|r|r|} 
    \hline
Passbands (\AA)  & loop 1 & loop 2 & loop 3 & loop 4 & loop 5 & loop 6 & loop 7 & loop 8 \\
\hline
HMI continuum   &30, 41  & 27 & 24, 31 & 22, 35 & 27, 35 & 22, 30 & 19, 27 & 24, 30\\ \hline
HMI Dopplergram & 27, 41 & 40 & 19, 23 & 24, 35 & 17, 26 & 20, 30 & 27, 35 & 24, 30 \\ \hline
    \end{tabular}
\end{table*}

\section{Discussion and Summary}
\label{sec:discussion}

To probe the source region of 3-min waves along the fan loops and coupling of the solar atmosphere, we utilised the property of frequency modulations of 3-min oscillations at different atmospheric heights. Such frequency modulations of 3-min waves are already noted in the transition region \citep[e.g.,][]{2001A&A...368..639F}, and corona \citep[e.g.,][]{2012A&A...539A..23S}. We explored this property of 3-min waves from the photosphere to the corona along the fan loop locations identified at different heights (see Figure~\ref{fig:fm}). Moreover, \citet{2015ApJ...812L..15K} noted similarity in periods of amplitude modulations in the range of 20-27 minutes across all atmospheric heights from the photosphere to the corona for 3-min and 5-min oscillations, which led them to conclude that photospheric p-modes externally drive 3-min slow magnetoacoustic waves observed in the coronal fan loops. Recently, \citetalias{2023MNRAS.tmp.2323R} performed detailed investigations on 3-min and 5-min oscillations from the photosphere to corona, and noted several amplitude modulation periods in the range of 9-14 min, 20-24 min, and 30-40 min. Their results revealed that 3-min waves observed in the coronal fan loops are driven by 3-min oscillations observed at the photospheric foot-points of these fan loops. 

In this work, we utilised less explored frequency modulations of 3-min oscillations, and found modulation periods in the range of 14-20 and 24-35 min throughout the photosphere to corona along fan loops. These modulations are correlated to each other at different atmospheric heights. Therefore, based on our findings of similar modulation periods of 3-min waves at all atmospheric heights, we conclude that 3-min waves in the upper atmosphere are direct result of 3-min oscillations observed at the photospheric umbral region similar to the findings of \citetalias{2023MNRAS.tmp.2323R}. In the solar atmosphere, density difference at different heights act as a barrier due to which a fraction of waves gets reflected back. Also due to decrease in density, steepening of amplitude takes place which results in shock formation in the lower atmosphere. This also affects the correlation values of frequency modulations between those heights and are clearly observed in correlations for passband pairs where sharp change in densities are expected. Both of these results strengthens our claim that 3-min waves observed in the coronal fan loops are driven by 3-min oscillations observed at the photospheric foot-points of these fan loops within the umbra and supports the models of \citet{1991A&A...250..235F} and \citet{2006ApJ...640.1153C}.

Contrary view exists on any connection between 3-min and 5-min oscillations observed in the solar atmosphere \citep[e.g.,][]{2015ApJ...812L..15K, 2017ApJ...836...18C}. Therefore, we also attempted to explore any coupling between 3-min and 5-min oscillations at the photosphere using HMI continuum and Dopplergram data. We found a reasonable correlation between modulations of 3-min and 5-min oscillations with HMI continuum data but found very poor correlation with HMI Dopplergram data (see Figure~\ref{fig:fm5}(c)). Therefore, any connection between 3-min and 5-min oscillations at the photosphere is still ambiguous and is a topic of future interest and exploration. 

Moreover, during the analysis, we have also noted that in comparison to amplitude modulation, frequency modulation is a more direct method to probe the connectivity of the solar atmosphere, especially to probe the connection between 3-min and 5-min oscillations at the photosphere as this relation or drift can be directly observed and analyzed with the wavelet tools. However, we need better frequency and spatial resolution data to explore these connections.

 In summary, we probed the magnetic coupling of the solar atmosphere using frequency modulations of 3-min waves observed from the photosphere to the corona.  These 3-min waves showed periodic modulations in their frequency with periods of about 14-20 min and 24-35 min at all the heights, and are correlated at different atmospheric heights except at photosphere. Results reveal that 3-min waves observed in the coronal fan loops are driven by 3-min oscillations observed at the photospheric footpoints of these fan loops in the umbral region, and thus, demonstrate magnetic coupling of the whole solar atmosphere.
 
\begin{acknowledgments}
The AIA and HMI data used here are courtesy of NASA/SDO and the AIA and HMI consortia. IRIS is a NASA small explorer mission developed and operated by LMSAL with mission operations executed at NASA Ames Research center and major contributions to downlink communications funded by the Norwegian Space Center (NSC, Norway) through an ESA PRODEX contract. Facilities: SDO (AIA, HMI), IRIS.
\end{acknowledgments}

\begin{furtherinformation}

\begin{orcids}
\orcid{0009-0005-9936-9928}{Ananya}{Rawat}
\orcid{0000-0002-0437-6107}{Girjesh}{Gupta}

\end{orcids}

\begin{authorcontributions}

AR performed data analysis and obtained figures. GG conceptualised the study and performed data correction. Both the authors contributed in paper writing. GG supervised the project.

\end{authorcontributions}

\begin{conflictsofinterest}
The authors declare no conflict of interest.
\end{conflictsofinterest}
\end{furtherinformation}

\end{document}